# Magnetocaloric effect and magnetic phase diagram of EuRhAl$_4$Si$_2$

Arvind Maurya[1], A. Thamizhavel[1], P. Bonville[2], and S. K. Dhar[1,*]

[1]Department of Condensed Matter Physics and Materials Science, Tata Institute of Fundamental Research, Homi Bhabha Road, Colaba, Mumbai 400 005, India
[2]CEA, Centre d'Etudes de Saclay, DSM/IRAMIS/Service de Physique de l'Etat Condensè and CNRS UMR 3680, 91191 Gif-sur-Yvette, France

## Abstract

EuRhAl$_4$Si$_2$, crystallizes in tetragonal crystal structure and orders antiferromagnetically at ~12 K. The isothermal magnetization along the two principle directions is highly anisotropic despite Eu$^{2+}$ being an *S*-state ion. The variation of entropy change, which is a measure of *MCE*, with field and temperature, calculated from the isothermal magnetization data taken at various temperatures along the principal crystallographic directions present interesting behavior in EuRhAl$_4$Si$_2$. In the magnetically ordered state the entropy change is non-monotonic below spin flip fields; however, in the paramagnetic region, it is negative irrespective of the strength of applied magnetic field. For *H* || [001] the maximum entropy change at 7 T is -21 J/Kg K around $T_\mathrm{N}$, which is large and comparable to the largest known values in this temperature range. The variation of the *MCE* with field strongly depends upon the direction of the applied magnetic field. Magnetic phase diagram of EuRhAl$_4$Si$_2$ derived from *M(H)* data is also constructed.

## Introduction

The Magnetocaloric effect (*MCE*) is one of the important aspects in anisotropic magnets with large spin, to explore basic physics as well as their application potential. *MCE* is usually observed as an isothermal magnetic entropy change ($\Delta S_M$) or an adiabatic temperature change ($\Delta T_{ad}$) when the material is subjected to a varying magnetic field. Large values of $\Delta S_M$ and $\Delta T_{ad}$ are considered to be important requirement for application potential of *MCE* materials. It is generally observed that a large *MCE* is observed in a material that undergoes first order magnetic phase transition [1, 2]. Recently, a series of quaternary compounds $RTAl_4Si_2$ (R = Rare-earth, T = Transition metal), crystallizing in an ordered variant of $KCu_4S_3$-type tetragonal unit cell, have been reported [3, 4]. Eu-analog of this series is found to exhibit highly anisotropic behavior, even in absence of total orbital angular momentum in its divalent state, which is primarily attributed to crystal electric field anisotropy in local moment magnetic systems. In these materials the ground state is speculatively antiferromagnetic with a small ferromagnetic component along the easy axis of magnetization. At intermediate fields an up-up-down sequence of *ab*-plane ferromagnetic blocks stacked along **c**-axis was suggested [5]. The isothermal magnetization data on $EuRhAl_4Si_2$ exhibit sharp first-order metamagnetic transition at relatively low fields in [001] direction. In view of this first order metamagnetic jump we investigate the *MCE* in the single crystalline sample of $EuRhAl_4Si_2$.

## Experimental

The single crystals of $EuRhAl_4Si_2$ were grown by self-flux technique with excess Al-Si flux as described in Ref. 3. Large shiny single crystals having platelet morphology with the flat plane corresponding to the (001) plane were obtained after centrifuging the flux. For the purpose of anisotropic magnetic measurements, the crystals were oriented along the principal crystallographic directions *viz.*, [100] and [001] by means of back reflection Laue method using a Huber Laue diffractometer. The oriented crystal was then cut appropriately by a spark erosion cutting machine. The magnetization data were taken in a Quantum Design superconducting quantum interference device (SQUID) magnetometer and vibrating sample magnetometer (VSM).

The *MCE* is typically calculated from the isothermal magnetization data at different temperatures or heat capacity at different fields. We have utilized isothermal magnetization data taken in zero field cooled state (Fig. 1) to calculate *MCE* by using the following relation.

$$\Delta S_M(0 \rightarrow H) = \int_0^H \left(\frac{dM}{dT}\right)_H dH$$

To make numerical calculations viable, the isothermal magnetization data at fixed intervals of magnetic fields (0.05 T) was obtained by 1401-point interpolation of 0-7 T experimental data. The matrix so obtained was transposed to obtain $M(T)$ data at different constant values of fields, which was differentiated numerically by finite difference method to get integrand $(dM/dT)_H$. Finally, entropy change ($\Delta S_M$) was derived by integrating the expression from zero to desired values of magnetic fields.

## Results and Discussion

From the magnetic susceptibility measurement, we found that EuRhAl$_4$Si$_2$ undergoes a first order transition from a paramagnetic to an incommensurate antiferromagnetic state at 11.7 K followed by a lock-in transition to a commensurate antiferromagnetic phase at 10.4 K [5]. This type of cascading transition in Eu compounds is not so unusual as it is seen in several other compounds [6-9]. The isothermal magnetization, along the [100] and [001] directions, measured at various fixed temperature for fields up to 7 T is shown in Figs. 1(a) and (b), respectively. For $H \parallel [100]$ the magnetization at 2 K increases with a small negative curvature and exhibit a spin re-orientation or a spin-flip process at around 5 T and then exhibit a saturation behaviour. As the temperature is increased this spin-flip field shifts to lower fields and finally vanishes above the ordering temperature. On the other hand, when the field is applied along the [001] direction at 2 K, the magnetization increases sharply even for small fields to one third of the saturation magnetization and exhibit a plateau up to a field of 1.8 T. At this point there is a sharp rise in the magnetization leading to a saturation value of ~7 µ$_B$/Eu at a magnetic field of 2T. The saturation magnetization value is in agreement to divalent

Eu-ion with $g_J$ = 2 and $S$ = 7/2. As observed for $H \parallel [100]$, along the [001] direction too, when the temperature is increased the spin-flip field at 2 T, shifts towards lower fields and gets smoothened as the magnetic ordering temperature is approached. Beyond 12 K, the magnetization does not show any anomaly.

The magnetization along [00] at 2 K possess a hysteresis at the sharp transition near 1.8 T, when the applied magnetic field was swept back to zero [5]. This signals the first order nature of the metamagnetic transition. Arrot plots of EuRhAl$_4$Si$_2$ derived from the zero field cooled virgin $M$ vs. $H$ curves measured at various temperatures are shown in Fig. 2, for the two principal crystallographic directions. It is obvious from the figure that the slope of the Arrot plot is negative below the magnetic ordering temperature; further confirming the first order nature as per the Banerjee criterion [10].

The *MCE* for selected temperatures and fields are shown in Fig. 3. *MCE*, defined as negative of the entropy change is mostly negative at low temperatures. For $H \parallel [100]$, it exhibits a peak in negative *MCE* quadrant. The height of the peak is maximum for 4-5 T data, which is close to spin flip field along the same direction. There is a crossover from negative to positive value of *MCE* at around 10 K, followed by a positive peak at ~12-13 K. The maximum observed value of *MCE* at 7 T for $H \parallel [100]$ and [001] amounts to 16 and 21 J/Kg K, respectively, revealing a significant magnetic entropy change in these systems. The *MCE* vs. temperature plots for $H \parallel [001]$ are similar to [100], except there is no apparent negative peak for $H \parallel [001]$ data. The maximum values of $\Delta S_M$ for a field change of 2 and 7 T are 6.5 J/Kg K and 21.1 J/Kg K, respectively in $H \parallel [001]$. It is of interest to mention that some of the prominent rare earth di-aluminides (Er, Dy)Al$_2$, which are promising magnetic refrigerants also possess similar value of $\Delta S_M$ in the temperature range 13 – 60 K [11]. The negative dip in the $\Delta S_M$ below 11 K, is due to the presence of antiferromagnetic component.

Fig. 4 represents the *MCE* data as a function of magnetic field at selected temperatures, which again follows closely to the magnetization data. There are clear anomalies at the spin flip/spin flop fields. For $H \parallel [100]$, the MCE at lower temperature (up to 4 K) is initially positive, before crossing the abscissa, exhibits a peak in negative *MCE* quadrant nearly at the spin flip field, after which it starts increasing towards positive values. 5-10 K data are qualitatively similar to that of lower temperatures except low field values are negative (but close to zero) even at lowest fields. Higher temperature traces show a monotonic variation. For $H \parallel [001]$, there is again highly nonlinear and non-monotonic variation of *MCE* is observed below 12 K. At 2 K, there is a wide inverse *MCE* plateau and after a negative step it does not vary much with further increment of field. At higher temperatures, there is a tendency to increase in *MCE* after spin flop.

To reveal the variation of *MCE* in *H-T* space, we have plotted the interpolated colour map of *MCE* in Fig. 5. The solid lines are demarcation for sign change in the *MCE*. The *H-T* region in the magnetically ordered state is divided into positive and negative *MCE* areas in a complex manner, which are tentatively due to the relative dominance of ferromagnetic and antiferromagnetic nature of the coupling between $Eu^{2+}$ spins resulting by various temperature and field dependent competing interactions. Microscopic studies are intended to understand the origin of such a complex *MCE* behavior below $T_N$.

Figure 5(a) shows the occurrence of hysteresis in magnetization loop of $EuRhAl_4Si_2$ along the **c**-axis. A closer view of the apparently flat plateau region in the high resolution magnetization data shows a couple of hysteretic tiny steps in addition to the large metamagnetic jump bearing the saturation value of the magnetization. These small steps are of height smaller than 0.1 $\mu_B$, and hence are susceptible to be ignored at first sight; accounting them under domain effects or spurious phenomena. They are also well captured in the derivative plot (Fig. 6b). A mapping of all the steps occurring in the magnetization data result in a phase diagram represented in Fig. 6c. The way, in which tiny steps feature a regular pattern in the phase diagram shows that they might be appearing due to some intricate mechanism. Interestingly, all the observed hysteresis in *M(H)* along [001] vanish above 5K. However, the phase diagram of $EuRhAl_4Si_2$ along the [100]-direction is typical for an antiferromagnetic material (Fig. 6d).

## Conclusion

In conclusion, we have grown the single crystal of $EuRhAl_4Si_2$ and studied the anisotropy in the *MCE*. This compound exhibits a cascading magnetic ordering at around 12 K. The first order nature of the magnetic transition is confirmed by means of the Arrot plots. The estimation of the *MCE* from the isothermal magnetization indicates a reasonably large value for $\Delta S_M$ compared to other prominent magnetocaloric materials. A sign reversal in the *MCE* is observed in *H-T* phase space; which is attributed to competing interactions at low temperatures.

The magnetic phase diagram of EuRhAl$_4$Si$_2$, particularly along [001] is interestingly exotic. The low temperature phase is hysteretic in nature. There are tiny steps in apparently flat plateau region in the *M(H)* along the **c**-axis in EuRhAl$_4$Si$_2$. However, their imprint producing a regular pattern in the phase diagram in *S*-state ion appears nontrivial and a further investigation into it is required to understand their origin.

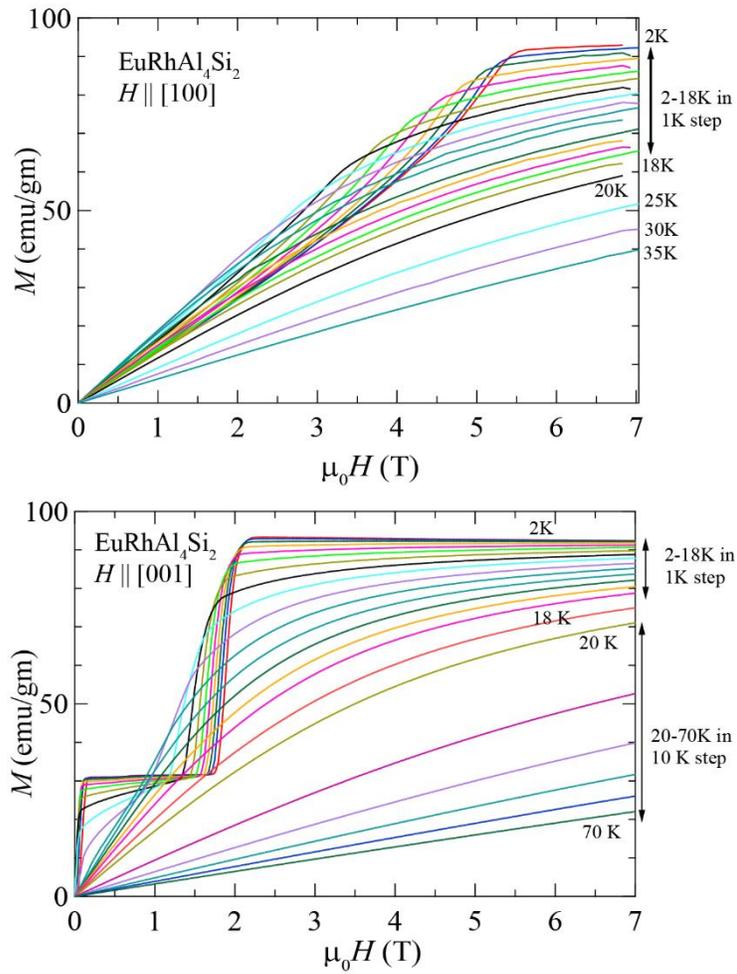

Figure 1. Isothermal magnetization data at selected temperatures of EuRhAl$_4$Si$_2$ when field is parallel to (a) **a** and (b) **c**-axes.

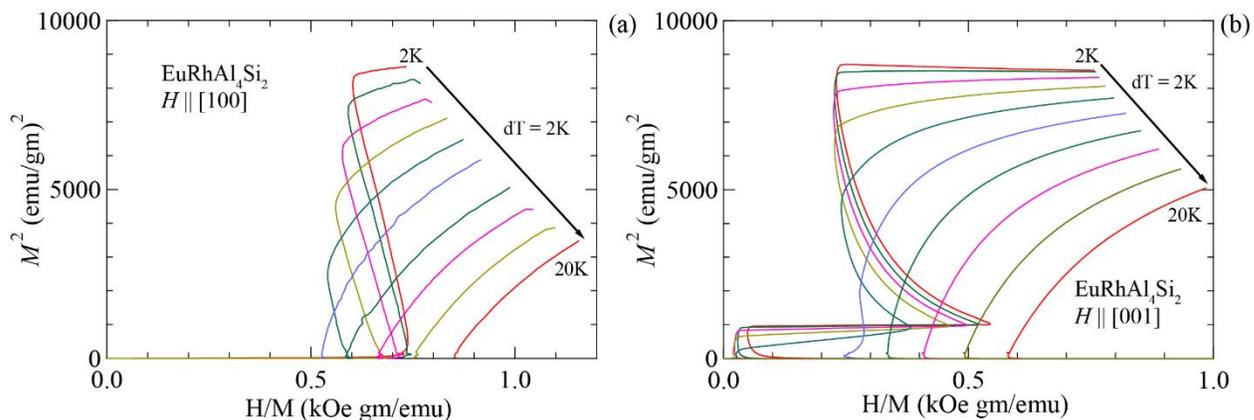
Figure 2. Arrot plots at selected temperatures of EuRhAl$_4$Si$_2$ when field is parallel to (a) [100] and (b) [001] directions.

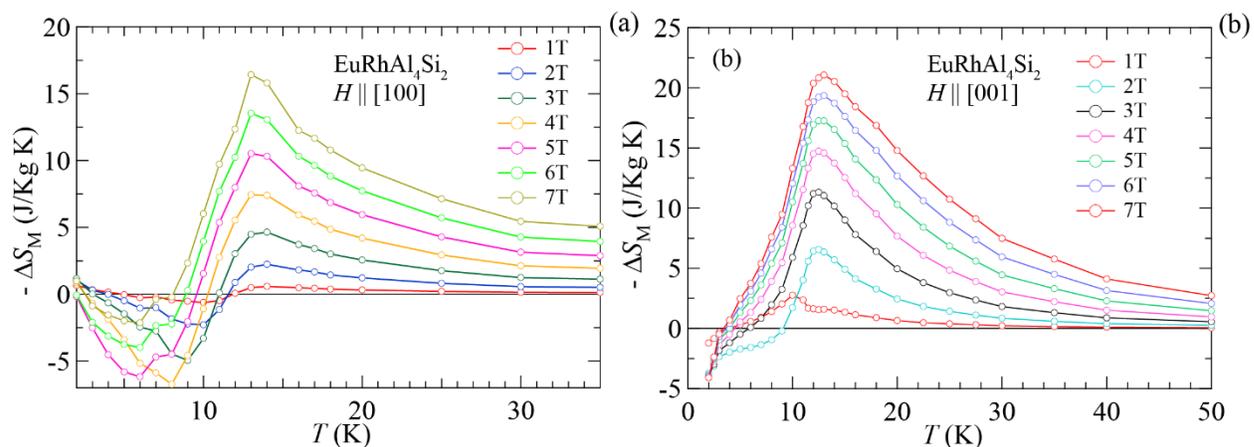
Figure 3. *MCE* in EuRhAl$_4$Si$_2$ as a function of temperature at selected magnetic fields for (a) $H \parallel [100]$ and (b) $H \parallel [001]$.

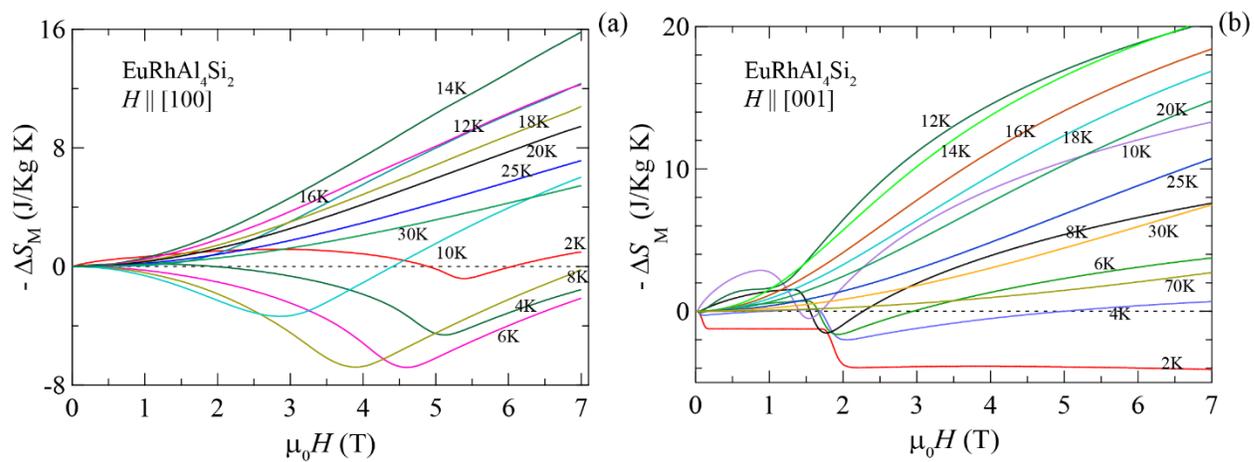

Figure 4. Evolution of *MCE* in EuRhAl$_4$Si$_2$ as a function of magnetic field at selected temperatures for (a) $H \parallel [100]$ and (b) $H \parallel [001]$.

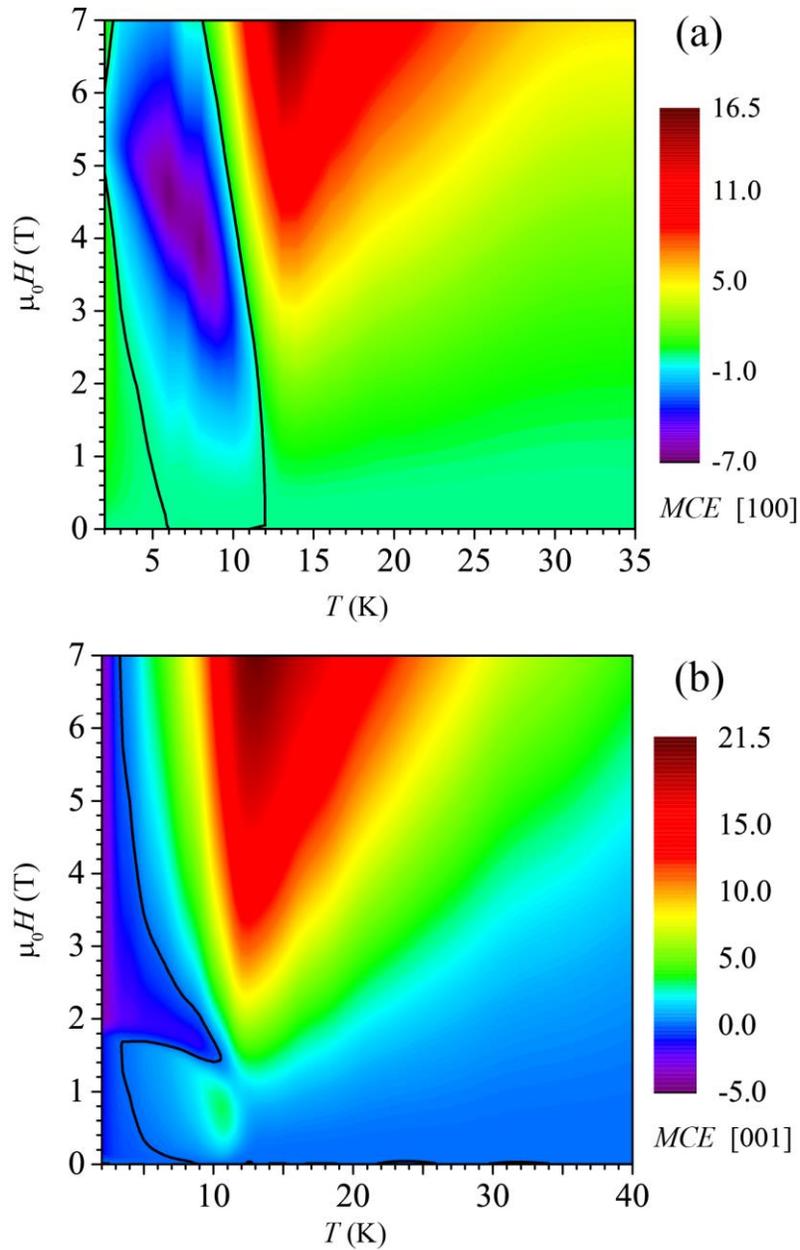

Figure 5. Colour mapped contour plot of *MCE*(*H*, *T*) in EuRhAl$_4$Si$_2$ for (a) *H* || [100] and (b) *H* || [001]. Solid lines separate the negative and positive values of *MCE*.

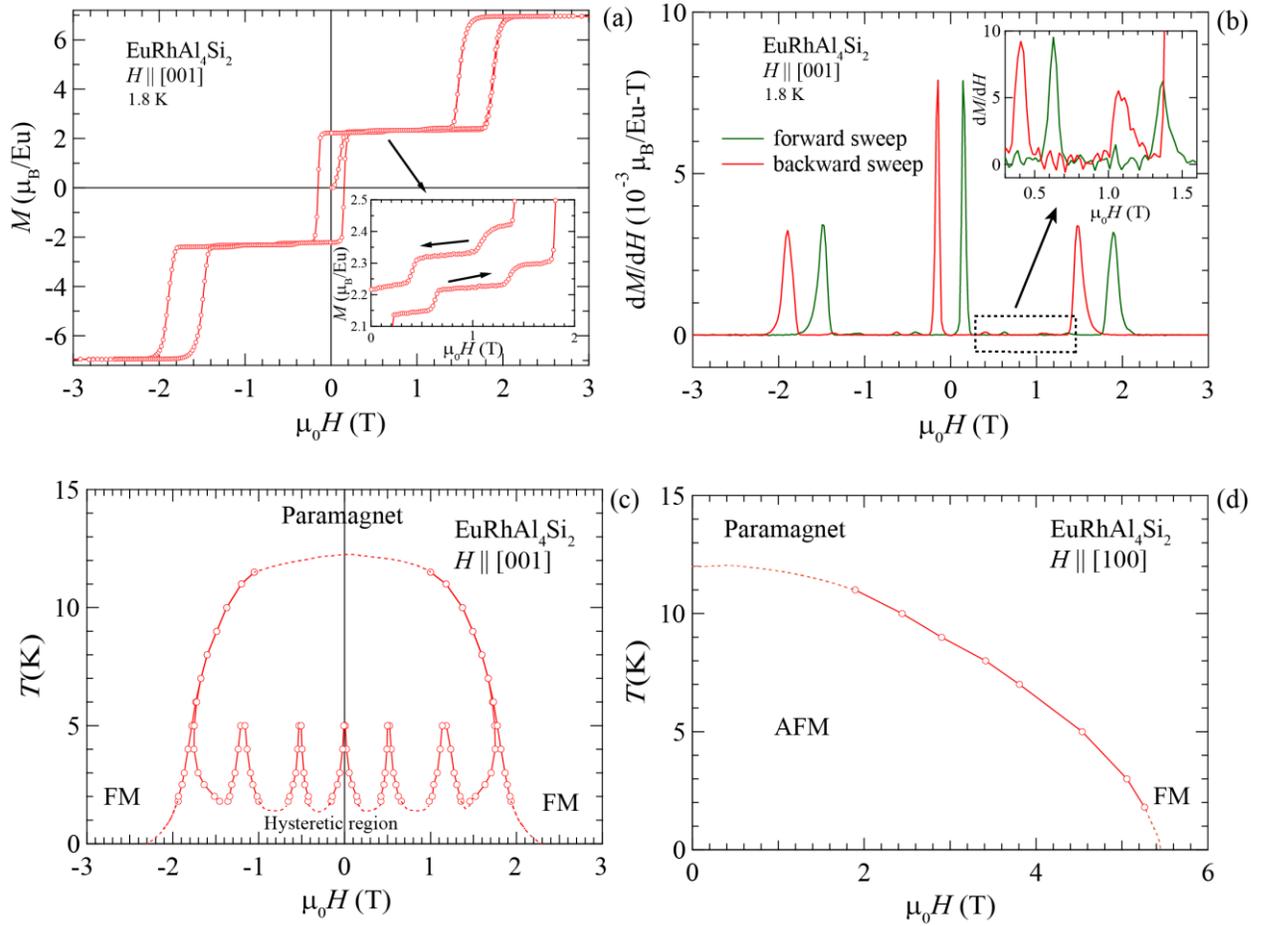

Figure 6. (a) Magnetization loop of EuRhAl$_4$Si$_4$ at 1.8 K for $H \parallel [001]$; the inset shows the expanded view of the plateau region. (b) d$M$/d$H$ vs. $H$ at 1.8 K in forward (green) and backward (red) field sweep; inset zooms up the small peaks between the major peaks. Magnetic phase diagram of EuRhAl$_4$Si$_4$ for (c) $H \parallel [001]$ and (d) $H \parallel [100]$. Solid lines are guide to the eyes and dotted lines are extrapolation.